\documentclass[namedreferences]{solarphysics}

\usepackage[hyperref,optionalrh,showbiblabels]{spr-sola-addons} 
\usepackage{graphicx}        
\usepackage{color}           
\usepackage{breakurl}        

\def\BB{{\bf {B}}}
\def\JJ{{\bf {J}}}
\def\vv{{\bf {v}}}
\def\xx{{\bf {x}}}
\def\XX{{\bf {X}}}
\def\ee{{\bf {e}}}

\begin{document}

\begin{article}
\begin{opening}

\title{Why are flare ribbons associated with the spines of magnetic null points generically elongated?}


\author[addressref=aff1,corref,email={d.i.pontin@dundee.ac.uk}]{\inits{D.I.}\fnm{David}~\lnm{Pontin}\orcid{0000-0002-1089-9270}}
\author[addressref=aff2,email={kg@nbi.ku.dk}]{\inits{K.}\fnm{Klaus}~\lnm{Galsgaard}\orcid{0000-0001-8882-1708}}
\author[addressref=aff3,email={Pascal.Demoulin@obspm.fr}]{\inits{P.}\fnm{Pascal}~\lnm{D{\' e}moulin}\orcid{0000-0001-8215-6532}}

\address[id=aff1]{School of Science \& Engineering, University of Dundee, Dundee, DD1 4HN, UK}
\address[id=aff2]{Niels Bohr Institute, Geological Museum, {\O}stervoldgade 5-7, 1350 Copenhagen K, Denmark}
\address[id=aff3]{Observatoire de Paris, LESIA, UMR 8109 (CNRS), F-92195 Meudon Principal Cedex, France}

\runningauthor{Pontin et al.}
\runningtitle{Flare ribbons in configurations with magnetic null points}

\begin{abstract}
Coronal magnetic null points exist in abundance as demonstrated by extrapolations of the coronal field, and have been inferred to be important for a broad range of energetic events. These null points and their associated separatrix and spine field lines represent discontinuities of the field line mapping, making them preferential locations for reconnection in the corona. This field line mapping also exhibits strong gradients adjacent to the separatrix (fan) and spine field lines, that can be analysed  using the {\it squashing factor}, $Q$. In this paper we make a detailed analysis of the distribution of $Q$ in the presence of magnetic nulls. 
While $Q$ is formally infinite on both the spine and fan of the null, the decay of $Q$ away from these structures is shown in general to depend strongly on the null point structure. For the generic case of a non-radially-symmetric null, $Q$ decays most slowly away from the spine/fan in the direction in which $|\BB|$ increases most slowly. 
In particular, this demonstrates that the extended, elliptical high-$Q$ halo around the spine footpoints observed by \citeauthor{masson2009} ({\it Astrophys.~J.}, \textbf{700}, 559, 
\citeyear{masson2009}) is  a generic feature.
The asymmetry of the halo of $Q$ contours around the spine/fan is shown to be strongest for the highest $Q$ contours, and increases as the null point asymmetry increases. This extension of the $Q$ halos around the spine/fan footpoints is in general important for diagnosing the regions of the photosphere that are magnetically connected to any current layer that forms at the null. In light of this, we discuss the extent to which our results can be used to interpret the geometry of observed flare ribbons in `circular ribbon flares', in which typically a coronal null is implicated.
In summary, we conclude that both the physics in the vicinity of the null and how this is related to the extension of $Q$ away from the spine/fan can be used in tandem to  understand observational signatures of reconnection at coronal null points.
\end{abstract}

\keywords{Magnetic fields, Corona; Flares, Relation to Magnetic Field; Magnetic Reconnection, Observational Signatures}

\end{opening}



\section{Introduction}\label{introsec}
As new generations of solar telescopes allow ever more detailed views of the Sun's atmosphere, the link between magnetic topological structures and observed sites of energy release becomes increasingly apparent. The magnetic structure of the corona is highly complex over a broad range of scales, as a result of the complex array of magnetic polarities that appear in a continually evolving pattern on the photosphere. The magnetic flux from each polarity region on the photosphere generically connects to many other flux patches of opposite polarity. The structure of the associated coronal magnetic field can appear bewilderingly complex, but advances in theory, modelling and observations have allowed a characterisation of the key features of the 3D structure -- such as likely sites for dynamic events to take place. One particular tool for analysing the coronal field structure is the magnetic field line mapping between positive and negative polarity regions of the photosphere. In particular, field lines along which this mapping is discontinuous -- usually {\it separatrix surfaces} associated with {\it magnetic null points} -- or has strong gradients -- at {\it quasi-separatrix layers} (QSLs) -- are now known to be likely sites for current accumulation and energy dissipation. These structures are defined and discussed in the following section.

In the last 20 years or so, a wealth of observational evidence has accumulated for energy release at both magnetic null points and QSLs in the form of flares, jets, and bright points \citep[e.g.][]{demoulin1997,fletcher2001,mandrini2006,luoni2007,masson2009,liu2011,zhang2012}. In each of these cases, the magnetic field structure in the corona must be inferred by employing some extrapolation method that uses the observed photospheric field as a boundary condition. One particular recent focus has been to understand in flaring regions how the observed $H_\alpha$ flare ribbons map to the coronal magnetic field structure -- and what one can subsequently deduce about the flare energy release process. In configurations containing QSLs, the footprints of these QSLs have been shown to be co-located with observed ribbons \citep{demoulin1993,dalmasse2015,savcheva2015,savcheva2016}. In specific cases a magnetic null point is also present. It has a fan separatrix and spines which define the topology of the magnetic configuration, and since the vicinity of the null is a preferential site for current accumulation and reconnection, so the footpoints of the spine and fan structures are often where the flare ribbons are located.  While the fan surface footprint naturally defines elongated ribbons, the spine lines define locally compact regions, so that one would expect compact ribbons -- however they are also observed to be elongated. {A link is observed between this elongation and the squashing degree $Q$ surrounding the fan and spines \citep[e.g.][]{mandrini2006} -- the reason for this link is explored herein}. \cite{masson2009} observed a so-called circular flare ribbon associated with the footprint of the separatrix surface of a coronal null point -- and they noted again the elongation of the spine footpoint ribbons. Since this observation, a number of further studies have confirmed that these findings are generic \citep{wang2012b,yang2015,liu2015}.

In this paper we make more concrete the link between the null point magnetic field structure, the geometries of associated features in the field line mapping, and the expected locations of flare ribbons. This is done by analysing the field line mapping in the vicinity of generic 3D null point structures, and relating this to the known properties of current sheet formation and magnetic reconnection around these nulls. The paper is organised as follows. We start in Section \ref{backsec} by discussing some necessary background on 3D magnetic topology and reconnection. In Sections \ref{linsec} and \ref{globalsec} the field line mapping in a linear null configuration and coronal separatrix dome configuration are analysed, respectively. In Section \ref{mhdsec} the field line mapping is studied in the context of magnetic reconnection around the null point in MHD simulations. We end in Section \ref{discuss} with a discussion.

\section{Background}\label{backsec}
\subsection{Magnetic null points}
In this paper we deal with magnetic null points in the solar corona. Such coronal null points have been demonstrated to exist in abundance by various surveys  of coronal magnetic field extrapolations, both potential and force-free \citep{regnier2008,edwards2015,freed2015}. A magnetic null is simply a location in space at which the magnetic field strength is exactly zero, $\BB={\bf 0}$, and in three dimensions (3D) this condition is met generically only at isolated points. The magnetic field in the vicinity of the null is characterised by a pair of {\it spine} field lines that asymptotically approach (or recede from) the null and a {\it fan} surface within which field lines radiate away from (or approach) the null point -- see \cite{lau1990,parnell1996} for a full description, and Figure \ref{config} for a visualisation. The fan surface forms a separatrix surface in the field, distinguishing two volumes of magnetic flux within which the field line connectivity is topologically distinct.

The simplest generic coronal structure involving a magnetic null point  occurs when a `parasitic' polarity region on the photosphere is surrounded by polarity region(s) of the opposite sign (and greater total flux). In this configuration a magnetic null is located where the  field contributions from the two polarities cancel. The fan separatrix surface then forms a dome structure and separates flux connecting the dominant polarity to the parasitic polarity (beneath the dome) from that which connects from the dominant polarity to locations further away on the photosphere. The effect of reconnection in such a separatrix dome configuration has been considered by \cite{edmondson2010,pontin2013}. Null points and their associated separatrix surfaces also occur in many more complicated topological configurations involving multiple null points, separatrix surfaces, and separators (separatrix surface intersections) -- see for example \cite{platten2014}. However, here we restrict our analysis to a single null point, considering a separatrix dome configuration in Section \ref{globalsec}.

3D null points are {\it one} of the preferential sites for reconnection in the corona. This is because in the perfectly-conducting limit, singular current layers are known to form at the null when rather general perturbations are applied \citep{pontincraig2005}. Therefore no matter how small the dissipation, non-ideal processes will eventually become important as the field around the null point collapses. Note that in an equilibrium there should be zero current at the null, since in general a pressure gradient cannot balance the Lorentz force in the vicinity of the null \citep{parnell1997}. Any non-zero current at the null will in general lead to a Lorentz force that drives the null point to collapse to form a current sheet \citep{klapper1996,pontincraig2005,fuentes2012,craig2014}.
The resulting reconnection may take different forms, the most general mode of reconnection being {\it spine-fan reconnection} that is associated with transfer of magnetic flux across the separatrix surface -- this process permitting in principle the release of significant stored magnetic energy \citep{antiochos1996,pontinbhat2007a,pariat2009,pontin2013}. There are also two other modes of reconnection at 3D nulls -- {\it torsional-spine} and {\it torsional-fan reconnection}, that involve a rotational slippage of field lines around the spine but involve no flux transfer across the separatrix \citep{pontin2011b}.

\subsection{The squashing factor and quasi-separatrix layers}\label{qsec}
The principal reason why magnetic null points were first proposed as sites of current accumulation and therefore magnetic reconnection in 3D is that at the null the field line mapping is discontinuous. It is now well established that 3D reconnection may also occur in the absence of null points or separatrices, and in particular natural sites for the formation of intense current layers are regions in which the field line mapping exhibits strong gradients.
Analysis of these gradients is typically performed by evaluating the {\it (covariant) squashing degree}, defined for planar boundaries by
\begin{equation}\label{qeq}
Q=\frac{(\partial U/\partial u)^2+(\partial U/\partial v)^2+(\partial V/\partial u)^2+(\partial V/\partial v)^2}{|(\partial U/\partial u)(\partial V/\partial v)-(\partial U/\partial v)(\partial V/\partial u)|},
\end{equation}
where $u$ and $v$ are field line footpoints on the `launch' boundary, and $U$ and $V$ are the footpoint locations on the `target' boundary, see  \cite{titov2002}. The general expression for non-planar boundaries can be found in Equations (11-14) of \cite{titov2007}. Note that the denominator in Eq.~(\ref{qeq}) can also be represented by $B_n/B_n^\star$ where $B_n$ and $B_n^\star$ are the field components normal to the boundaries at the launch and target footpoints, respectively. Numerically it is usually more stable to use this expression in practice. One potential weakness of $Q$ as defined in Eq.~(\ref{qeq}) is that the values obtained depend on the orientation at which field lines intersect the launch and target boundaries. An alternative formulation is the {\it perpendicular covariant squashing factor}, $Q_\perp$, as defined in equations (30-36) of \cite{titov2007}, which removes such projection effects by evaluating the mapping deformation for infinitesimal perpendicular planes at the locations of the launch and target surfaces.

Bundles of magnetic field lines along which $Q$ or $Q_\perp$ have large values ($\gg2$, the minimum value, obtained for a uniform field) are known as {\it quasi-separatrix layers}, or {\it QSLs}. It has been demonstrated that these are natural locations for accumulation of intense currents, using both modelling approaches \citep{titov2003,galsgaard2003,aulanier2005,effenberger2011} and solar observations \citep[e.g.][]{demoulin1997,mandrini2006}.
The term QSL comes from the fact that true separatrices can be thought of as a limiting case of a QSL \cite[see][for a detailed exposition]{demoulin2006}. In particular, due to the  discontinuity in the field line mapping at a separatrix, $Q$ is by definition infinite there. In addition, $Q$ must also be large but finite in the region adjacent to the separatrix -- and on this we focus in Section \ref{linsec}.

\subsection{The nature of 3D reconnection}\label{recsec}
In order to understand energy release mediated by magnetic reconnection in 3D, it is important to understand a key property of 3D reconnection. Specifically, in contrast to the 2D case reconnection in 3D {\it always occurs in a finite volume}. That is, rather than field lines breaking and rejoining at a single point (the X-point) as in 2D, field lines change connectivity continuously throughout the (finite-sized) non-ideal region \citep{priest2003a}. This non-ideal region is in general any region within which the electric field component parallel to the magnetic field ($E_\|$) is non-zero, and for which $\int E_\|\, ds\neq 0$, the integral being evaluated along field lines \citep{schindler1988}.
A consequence of the breaking and rejoining of field lines throughout the non-ideal region is as follows. For reconnection in the absence of separatrices, for example at QSLs, there is an everywhere continuous `flipping' or  `slipping' of reconnecting field lines \citep{priest1995}. One can also distinguish this flipping motion further, to {\it slipping} or {\it slip-running}, depending on whether the velocity of the apparent field line motion is sub- or super-Alfv{\' e}nic, respectively \citep{aulanier2006}. Such slipping motions are now observed during energy release in the corona \citep{aulanier2007,sun2013,li2014,dudik2014}.

All of the above statements hold true for 3D null point reconnection. In particular, the reconnection happens not only at the null point itself, but {\it throughout the non-ideal region (current sheet) surrounding the null} \citep{pontin2004,pontinhornig2005}. Therefore what we call `null point reconnection' is more precisely reconnection that occurs within a finite region (the current layer) surrounding a null -- the importance of the null being that it is a favourable site for intense currents to develop (as described above). There is still a continuous change of field line connectivity within the current sheet, and an apparent flipping motion of field lines. However, in the presence of a null there is also one discontinuous jump of connectivity, for every field line that is reconnected through either the spine of the null or the separatrix (fan) surface.
For a given reconnection event, if there is a null point in the non-ideal region the field line velocity must by necessity be `slip-running' near the null, since it is infinite at the spine/fan. For the case of a single null point in a separatrix dome configuration, that we examine below, the expected patterns of the field line slippage motions were described in detail by \cite{pontin2013}.


\section{Linear null point}\label{linsec}
\subsection{Preliminaries}
In the following section we examine the distribution of the squashing degree, $Q$, in the vicinity of null points.
First we consider the simplest case of a linear null point. Note that any generic 3D null point can be represented locally (i.e.~sufficiently close to the null) by this linearisation  --  and conversely for the null point to be topologically stable, the linear term in the Taylor expansion of $\BB$ about the null point must be non-zero \citep{hornig1996}. (Topological stability implies that an arbitrary perturbation does not destroy the topology, in contrast to the case where the first non-zero term is the quadratic term -- those higher order nulls are topologically unstable since an arbitrary perturbation of $\BB$ changes the topology.) Note that there are two factors that influence the variation of $Q$ in the vicinity of a null point and associated separatrix: one is that $Q$ is formally infinite for spine and fan field lines due to the discontinuity in the field line mapping. There is then a characteristic decay of $Q$ away from these field lines. The other factor is that field lines in the fan surface typically become oriented parallel to one of the fan eigenvectors (corresponding to the largest fan eigenvalue) at larger distances. This naturally means there is a stronger divergence of field lines away from the fan eigenvector direction corresponding to the smaller fan eigenvalue. This leads to a rotational asymmetry of the decay of $Q$ away from the spine and fan -- as demonstrated below.

\begin{figure}[t]
\centering
(a)\includegraphics[width=0.45\textwidth]{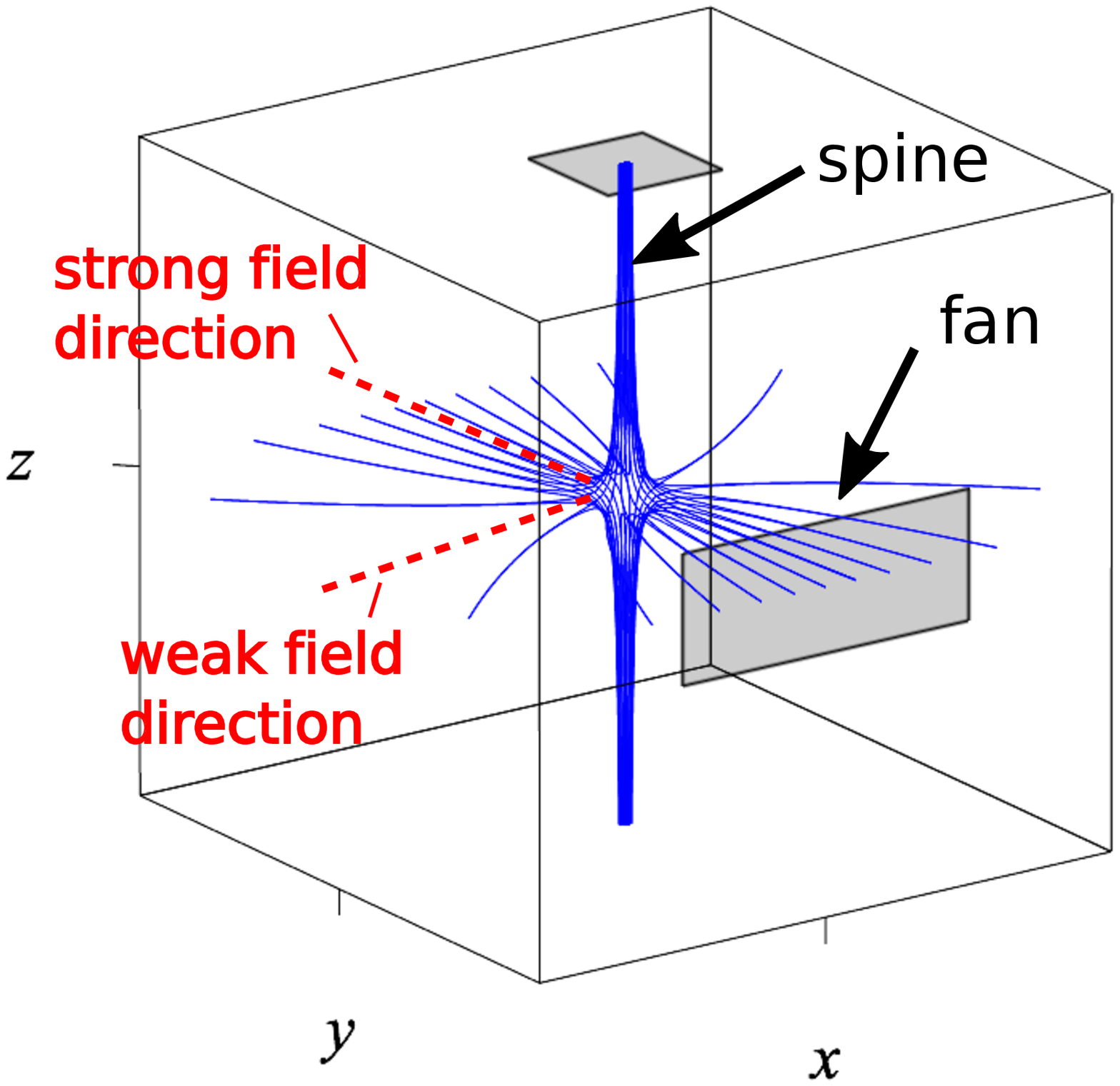}
(b)\includegraphics[width=0.45\textwidth]{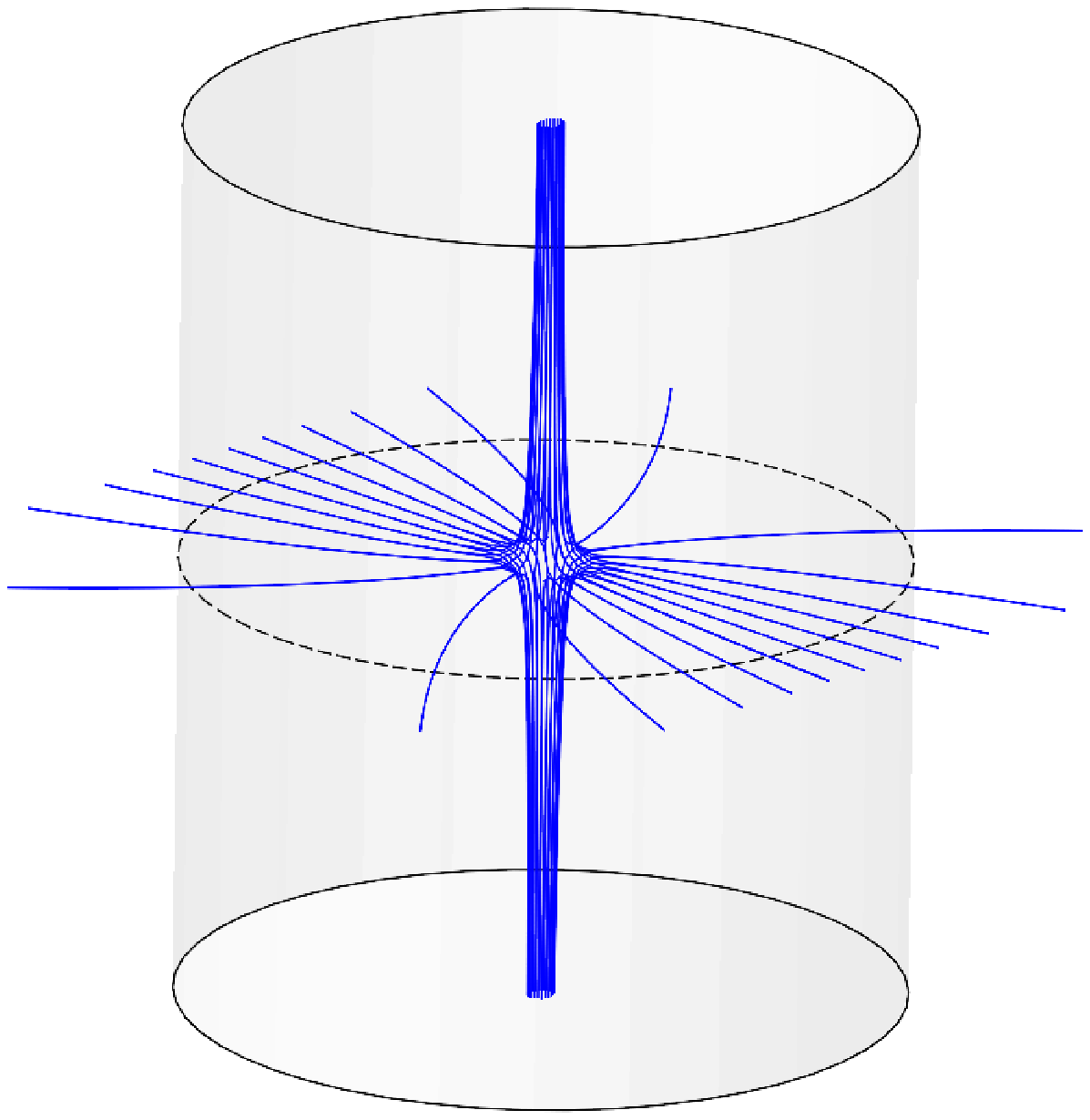}
\caption{Boundary surfaces used for the calculation of the squashing factor $Q$ or $Q_\perp$ in (a) Section \ref{planar} and (b) Section \ref{cylinder}. Field lines are plotted for $k=0.4$ ($k$ defines the asymmetry of the null-point field as defined by Eq.~\ref{beq}).}
 \label{config}
\end{figure}
Evaluation of the squashing factor requires that we select two surfaces that each field line intersects once and only once. We consider two cases: in the first case we take both boundaries to be planar (Figure \ref{config}a), and in the second case we take a plane of constant $z$ and a circular cylinder surface (see Figure \ref{config}b). When both surfaces are planar we can obtain exact expressions for $Q$ and $Q_\perp$, whereas for the cylindrical boundary we must evaluate them numerically.

\subsection{Squashing factor between two planar boundaries}\label{planar}
Consider an equilibrium magnetic null point (zero current).  The field can be represented by
\begin{equation}\label{beq}
\BB=B_0
\left(\begin{array}{ccc}
k & 0 & 0 \\ 0 & 1-k & 0 \\ 0 & 0 & -1
\end{array}\right)
\left(\begin{array}{c}
x\\ y\\ z
\end{array}\right)+ \mathcal{O}(r^2),
\end{equation}
with $0<k<1$, where we have chosen to orient the coordinate system such that the spine lies along the $z$-axis, the fan surface is coincident with the $z=0$ plane, and the two eigenvectors of $\nabla\BB$ are parallel to the $x$ and $y$ axes. The corresponding eigenvalues are unequal when $k\neq 1/2$, and in the $xy$-plane the field strength increases most quickly away from the origin along the direction of the eigenvector associated with the largest eigenvalue -- the $x$-direction for $1/2<k<1$. We refer below to this direction as the {\it strong field direction} in the fan, and correspondingly to the orthogonal direction as the {\it weak field direction}. {In the corona, one would in general expect field lines in the strong field direction to connect to the strongest nearby photospheric flux concentrations (though this need not necessarily be the case)}.
Taking two planar boundaries as shown in Figure \ref{config}(a), we can obtain a mathematical expression for how $Q$ decays away from (say) the spine along the strong and weak field directions in the fan. 

Here we take the `launch plane' for field lines to be $z=a$ and the target plane to be $x=b$ ($a$ and $b$ constant). That is, we consider the mapping ~$(x,y,a) \to (b, Y, Z)$. For simplicity we only examine how $Q$ decays from the spine along the $x$-axis (setting $Y=y=0$). This corresponds to the strong field direction in the fan if $1/2<k<1$, and the weak field direction for $0<k<1/2$. We examine the decay in arbitrary directions in the next section. 
Setting $y=Y=0$, we obtain the following expression for $Q$ as a function of $x$, the distance of the launch footpoint from the spine 
\begin{equation}\label{q_cart}
Q(y=0)=\frac{bk}{a}\left(\frac{x}{b}\right)^{(2-{2}/{k})}  + \frac{a}{bk} \left(\frac{x}{b}\right)^{-(2-{2}/{k})},
\end{equation}
{see Appendix \ref{app}}. We see that, as expected, $Q\to\infty$ as $x\to 0$ (the spine footpoint). We are interested in the behaviour for small $x$, and  since $0<k<1$, the relevant term close to the spine is the first one, so that
\begin{equation}
Q(y=0) \approx \frac{bk}{a}\left(\frac{x}{b}\right)^{(2-{2}/{k})}.
\end{equation}
This implies that in the weak field region ($0<k<1/2$) $Q$ should be larger (for fixed $x$) since we have a larger negative exponent. This is to be expected given the strong field line divergence in the $xy$-plane in this region. Hence a level curve of $Q$ would be expected to be elongated along the weak field direction.
Note that for the rotationally symmetric case ($k=1/2$) we have that $Q$ decays like $1/r^2$.

Consider now the distribution of $Q_\perp$ (see Section \ref{qsec}). Evaluating $Q_\perp$ on the $x$-axis as before we obtain the following expression
\begin{equation}\label{qperp_cart}
Q_\perp(y=0)=\frac{x^4k^2[(x/b)^{-2/k}b^2+a^2]+(x/b)^{2/k}b^4[a^2-k^2x^2]}{k^2b^2x^2\sqrt{k^2x^2+a^2}\sqrt{b^2+a^2(x/b)^{2/k}}}
\end{equation}
({Appendix \ref{app}}). For $x\ll b$, we have the same dominant scaling in $x$ as before, specifically
\begin{equation}\label{qp_app}
Q_\perp(y=0) \approx \frac{bk}{a}\left(\frac{x}{b}\right)^{2-2/k}.
\end{equation}
That the behaviour of $Q$ and $Q_\perp$ is identical in this plane is expected since in this plane as we get close to the spine and fan the field lines intersect the boundaries approximately perpendicular.

We now examine the dimensions of contours of $Q$ on the spine boundary ($z=\pm a$). Specifically, we rearrange Equation (\ref{q_cart}) to find the radius $x=r_x(Q_0)$ at which $Q_\perp=Q_0$. Assuming that $x/b\ll 1$ we can directly invert Equation (\ref{qp_app}) to obtain
\begin{equation}\label{rxeq}
r_x(Q_0)=b \left( \frac{aQ_0} {b}  \right)^{k/(2k-2)}
\end{equation}
One can obtain from this the same information along the $y$-axis by making the replacement $k\to 1-k$ to give $r_y(Q_0)$.
\begin{figure}[t]
\centering
\includegraphics[width=0.9\textwidth]{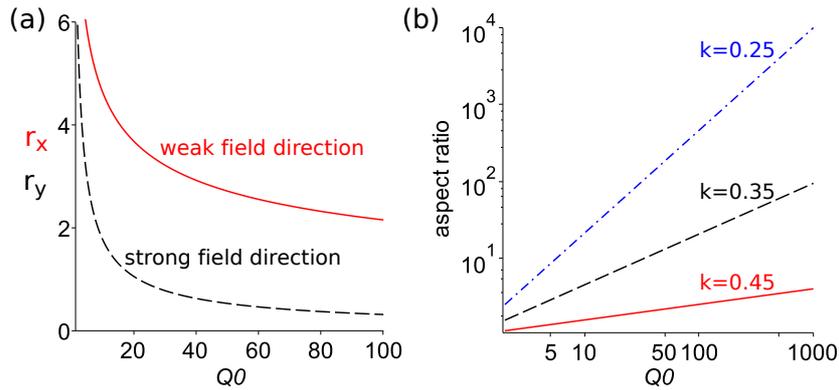}
\caption{(a) For fixed $k=0.4$, $r_x(Q_0)$ (red) and  $r_y(Q_0)$ (black dashed), as defined by Eq.~(\ref{rxeq});
(b) Log-log plot of the aspect ratio $r_x/r_y$ (see Eq.~\ref{aspectratio}), for $a=b$ and fixed $k=0.45$ (red), $k=0.35$ (black, dashed) and $k=0.25$ (blue, dot-dashed).
}
 \label{r_fn_q}
\end{figure}
As shown in Figure \ref{r_fn_q}(a), for $k=0.4$ the spacing of $Q_\perp$ contours drops off more slowly in the weak field direction, consistent with above. 

Now let us analyse the asymmetry in the $Q_\perp$ contour produced
by the asymmetry in null eigenvalues in more detail, by examining the relative decay of $Q_\perp$ from the spine footpoint along the two axes.  Using Eq.~(\ref{rxeq}) we have that 
\begin{equation}\label{aspectratio}
\frac{r_x(Q_0)}{r_y(Q_0)}=\left(\frac{aQ_0}{b}\right)^{(k-1/2)/(k(k-1))}.
\end{equation}
In Figure \ref{r_fn_q}(b) we plot the ratio $r_x/r_y$ for $a=b$ and three particular values of $k$. 
We see that, as expected, this aspect ratio increases as the magnetic field asymmetry increases. Furthermore, we observe that for high $Q_\perp$ values the $Q_\perp$ distribution is highly `non-circular' even for moderate values of $k$, while lower level contours of $Q_\perp$ only show high asymmetry when $k$ is far from 0.5 (note that Figure \ref{r_fn_q}(b) is a log-log plot). 
Note also that a full range of degrees of null point eigenvalue ratios (corresponding to $k\in (0,1)$) is obtained in solar extrapolations, and even for a moderate asymmetry (values of $k$ close to 0.5) the eccentricity is significant. {For example \cite{demoulin1994} analysed 6 nulls with $k=0.08$ -- $0.33$, the configuration analysed by \cite{masson2009} contained a null point with $k\approx 0.11$, and \cite{freed2015} carried out an extensive survey of potential field extrapolations from three years' worth of magnetogram data, identifying 1924 coronal null points -- choosing the orientation such that $0\leq k\leq 0.5$ their data shows a relatively uniform distribution of $k$ values between 0.1 and 0.5, with a mean value of $k$ of 0.26 and a standard deviation of 0.11.}

\subsection{Squashing factor for a surface encircling the fan}\label{cylinder}
The calculations of the previous section allow us to visualise the $Q$ distribution along the coordinate directions. However, since the vertical planar boundary intersects only a subset of the fan field lines, they do not allow us to visualise the $Q$ distribution all around the spine or fan footpoints. In order to do this we must choose a target plane that intersects all fan field lines (such that all field lines passing close to the null intersect both the launch and target surfaces).
\begin{figure}
\centering
\includegraphics[width=0.8\textwidth]{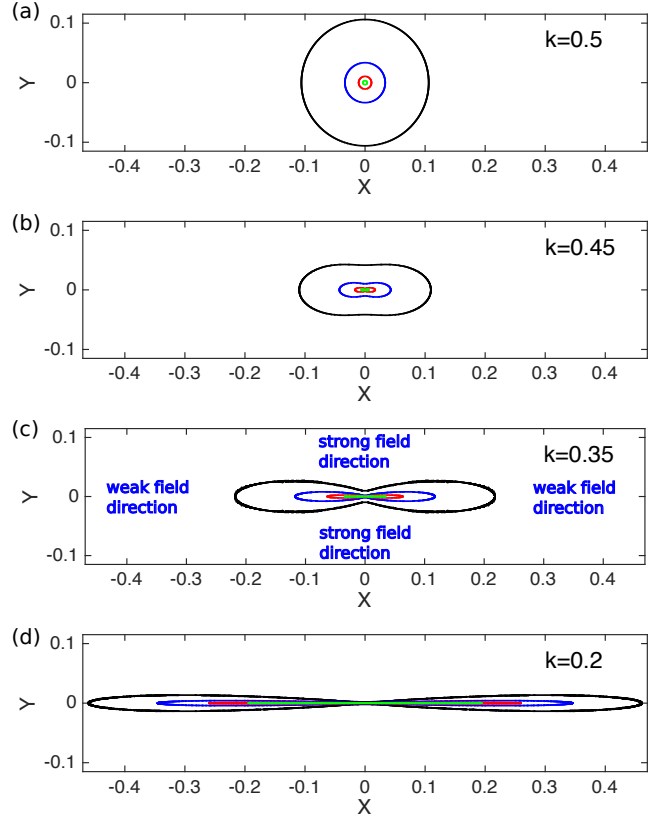}
\caption{Contours of $\log_{10} (Q_\perp)$ as calculated between surfaces $z=a=1$ and $r=b=1$ -- visualised on $z=1$ for different $k$. Contour levels are $\log_{10} (Q_\perp)=\{2,3,4,5\}$, and are coloured $\{$black, blue, red, green$\}$.}
 \label{qperp_contours}
\end{figure}
As such, we now calculate $Q$ between a planar `launch' surface at $z=a$  intersecting the spine, and a cylindrical `target' surface at $r=b$ encircling the fan, $a,b$ constant (Figure \ref{config}b). That is, we study the mapping generated by the field lines $(x,y,a)\to (b,\Theta, Z)$.
In this case we are unable to obtain a full analytical expression for the field line mapping and its inverse, and thus we evaluate $Q$ numerically.
We integrate between $10^5$ and $10^6$ field lines from a rectangular grid of starting points at $z=a$ to obtain their intersections with $r=b$, then perform derivatives of the mapping using a fourth-order-accurate centred difference over that grid. In this case we evaluate $Q_\perp$, since especially in the regions of strongly diverging field lines (the weak field region, around the $x$-axis for $k<1/2$) the field lines intersect the circular cylinder surface far from perpendicular. 

\begin{figure}[t]
\centering
\includegraphics[width=0.65\textwidth]{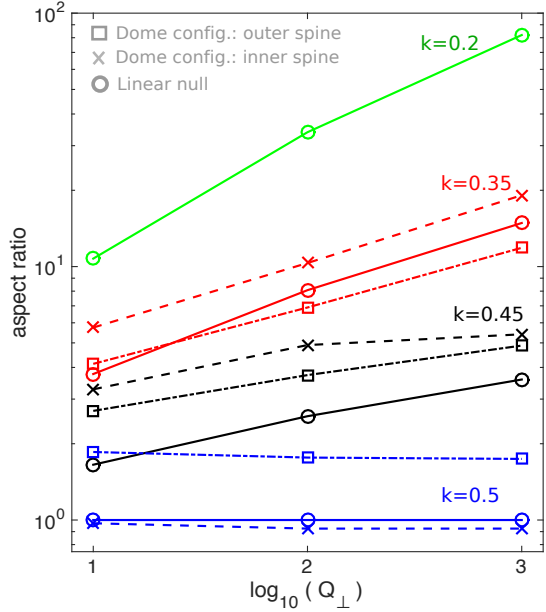}
\caption{Aspect ratio of $Q_\perp$ contours on the spine boundary,  defined as the maximum contour extent in the long direction ($x$ for the linear null) divided by the maximum contour extent perpendicular to that ($y$ for the linear null). Circles: $Q_\perp$ calculated between the surfaces $z=a=1$ and $r=b=1$ for the linear null point in Eq.~(\ref{beq}). Crosses, squares: $Q_\perp$ around the inner and outer spine footpoints, respectively, for the separatrix dome configuration in Eq.~(\ref{b_pot}). Blue, $k=0.5$; black, $k=0.45$; red, $k=0.35$; green, $k=0.2$.}
 \label{qperp_contours_aspect}
\end{figure}

The  $Q_\perp$ maps at $z=a$ for different $k$ are presented in Figure \ref{qperp_contours}. As predicted by the planar boundary analysis, we see a stretching of the $Q_\perp$ contours along the weak field direction in the fan ($x$ for $k<1/2$). We also see that the contours do not form simple ellipses, but are pinched in the middle so that their maximum extension is at finite $x$ values. 
Therefore in order to measure the asymmetry of the $Q_\perp$ distribution around the spine, it is arguably most useful to determine the largest extent of these contours along the $x$ and $y$ directions, rather than simply examining the profile along the coordinate axes. In Figure \ref{qperp_contours_aspect} we plot (with circles and solid lines) the contour aspect ratio defined as the maximum contour extent along $x$ divided by the maximum  extent along $y$ (over all $x$). 
The plot demonstrates the same trends as observed in the previous section, specifically that the aspect ratio is greater for both increased null point asymmetry (smaller $k$ for $k< 0.5$) and for higher $Q_\perp$ levels. Note that the qualitative features discussed above are present if one considers $Q$ instead of $Q_\perp$.

\begin{figure}
\centering
\includegraphics[width=0.7\textwidth]{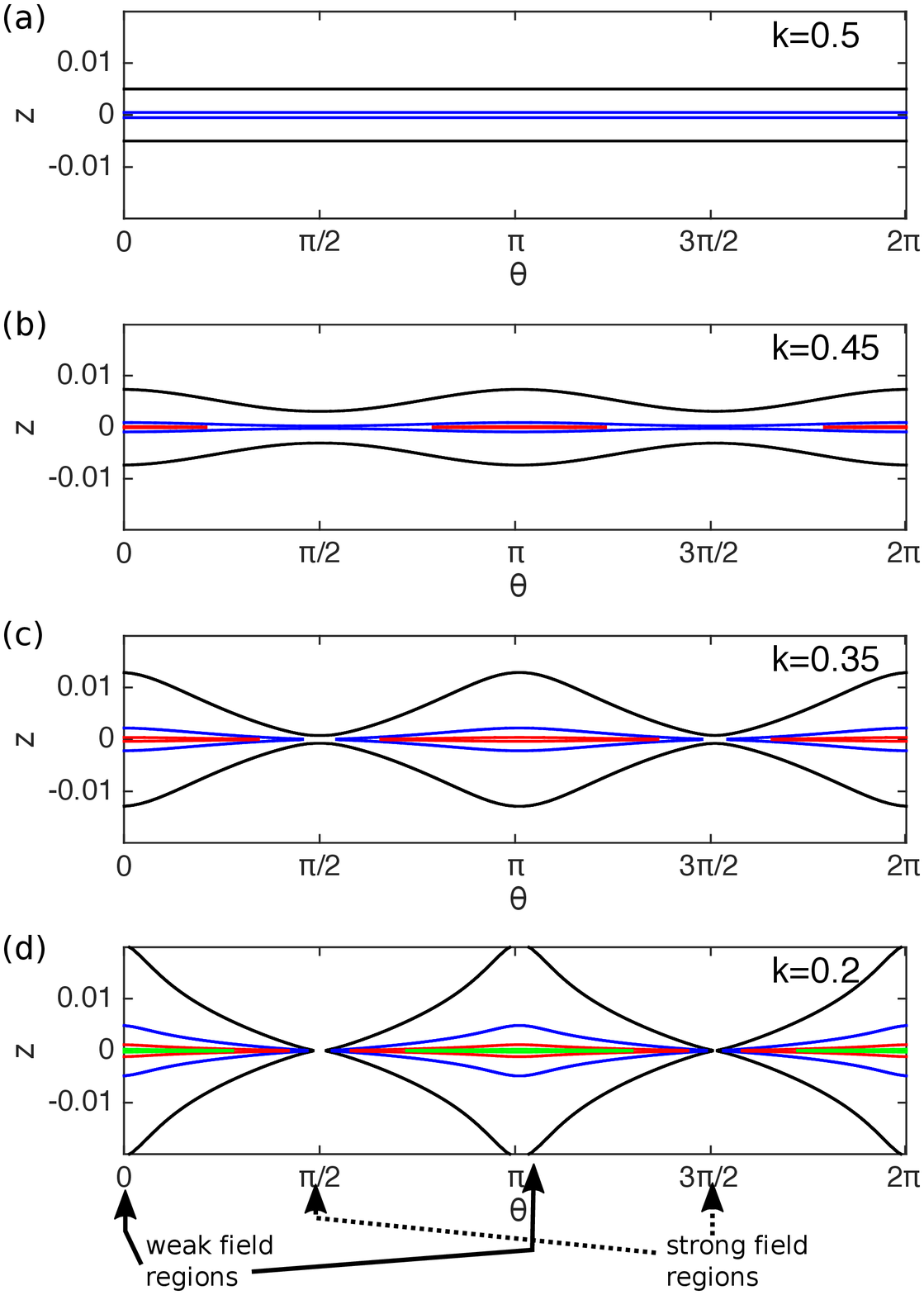}
\caption{Contours of $\log_{10} (Q_\perp)$ as calculated between surfaces $z=a=1$ and $r=b=1$ -- visualised on $r=1$ for different $k$. Contour levels are $\log_{10} (Q_\perp)=\{2,3,4,5\}$, and are coloured $\{$black, blue, red, green$\}$.}
 \label{qperp_contours_targetz}
\end{figure}
We can now perform the same $Q_\perp$ calculation procedure but with the launch and target boundaries reversed, in order to find the pattern of $Q_\perp$ in the vicinity of the fan surface -- shown in Figure \ref{qperp_contours_targetz}. We observe that $Q_\perp$ decreases in some locations and increases in others as we increase the asymmetry of the null. The weak field region is along the $x$-axis, which corresponds to $\theta=0,\pi$, by the usual convention. Again the widest $Q_\perp$ contours are located in the vicinity of these weak field directions. That is, for a given distance from the separatrix (height $z$), $Q_\perp$ is largest along the weak field direction.

The results above are entirely consistent with those of e.g.~\cite{masson2009}. In particular, we see that one obtains extended $Q_\perp$ (or $Q$) contours around the spine for even a moderate degree of null point asymmetry. Of course other global features of the field could well also contribute, but the figures show that the high $Q$ region in which the null is embedded is not a special additional feature of the particular field studied by \cite{masson2009}, but is rather a natural consequence of having a coronal null that is not rotationally symmetric.

\section{Current layer formation in MHD simulations, and flare ribbon locations}\label{mhdsec}
\subsection{Simulation setup and results}
In this section we ask the question: what is the relation between current layers that form at 3D nulls and the distribution of the squashing factor identified in the previous section, and to what extent can we expect the $Q$ (or $Q_\perp$) profile to predict where flare ribbons might be observed?
We consider resistive MHD simulations similar to those of \cite{galsgaard2011b}. Specifically, at $t=0$ in our simulations we have a linear magnetic null point of the form 
\begin{eqnarray}\label{beq:sim}
\BB &=& B_0 [
y(2k-1)\cos\phi\sin\phi + x(k\cos^2\phi + (1-k)\sin^2\phi),\\
&&y((1-k) \cos^2\phi + k\sin^2\phi) + x(2k-1)\cos\phi\sin\phi, -z], \nonumber
\end{eqnarray}
which reduces to Eq.~(\ref{beq}) for $\phi=0$. For $\phi\neq 0$, the fan plane eigenvectors are rotated by an angle $\phi$ with respect to the coordinate axes. The simulation domain is $x,y\in[-3,3],\, z\in[-0.5,0.5]$ and we set $B_0=1$. We apply a driving velocity on the domain boundaries that advects the spine footpoints in opposite directions on opposite $z$-boundaries, in the $y$ direction. Specifically, we let
\begin{eqnarray}
\label{driver.eq}
\vv(z=\pm0.5) & =  & \pm \ee_y \,v_0 \tanh\left(\frac{t}{T}\right)\,\left(\tanh\left(\frac{x-x_0}{x_h}\right)-\tanh\left(\frac{x+x_0}{x_h}\right)\right) \\
         & &             ~~~~~~ \times~  \left(\tanh\left(\frac{y-y_0}{y_h}\right)-\tanh\left(\frac{y+y_0}{y_h}\right)\,\right) , \nonumber
\end{eqnarray}
with $v_0=0.02, T=0.1, x_0=0.3,y_0=0.6,x_h=y_h=0.2$. Outside these regions $\vv={\bf 0}$ on all boundaries and $\BB$ is line-tied. {The plasma density and pressure are initially uniform, $\rho_0=1$, $p_0=0.0333$, so that the driving velocity is highly sub-sonic and sub-Alfv{\' e}nic. In addition, the resistivity $\eta=2\times10^{-3}$, also spatially uniform, throughout.}

As the simulation proceeds the stress injected by the boundary driving focusses around the null point -- the null point collapses and a current sheet forms around it \citep[see][]{pontinbhat2007a}. We examine three different simulations; in the first the null point is rotationally symmetric, while in the second and third we take an asymmetric null with $k=0.35$ corresponding to frame (c) in Figures \ref{qperp_contours} and \ref{qperp_contours_targetz}. Specifically, we use parameters in Eq.~(\ref{beq:sim}) as follows: {\it simulation 1}: $k=0.5,\phi=0$; {\it simulation 2}: $k=0.35, \phi=0$; {\it simulation 3}: $k=0.35, \phi=\pi/4$. Hereafter we analyse the state reached in the simulations at time $t=3$, this constituting a representative time by which the current layer has formed and reconnection is underway.
\begin{figure}[t]
\centering
\includegraphics[width=\textwidth]{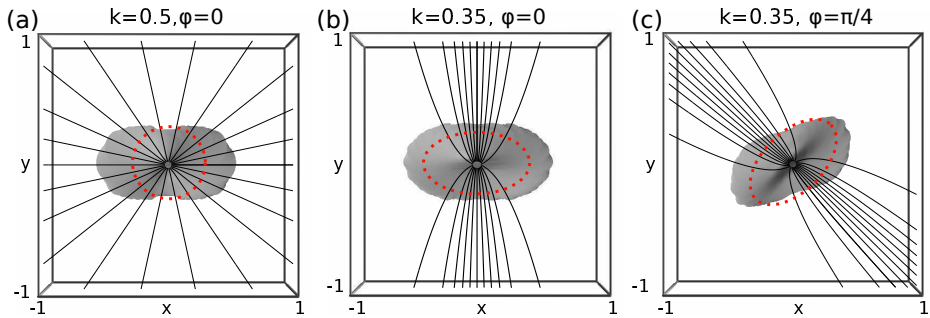}
\caption{Current density isosurface at level $|\JJ|=2.5$ at $t= 3.0$ from the MHD simulations (Section \ref{mhdsec}). The view is down onto the fan surface; overlayed are fan field lines. The dashed red curve indicates the dimensions of the $\beta=1$ curve in the $z=0$ plane.}
 \label{jsim}
\end{figure}
The current layer formed at $t=3$ is shown for each simulation in Fig.~\ref{jsim}. In each case the current is maximum at the null point. Note that the current layer extends from the high-$\beta$ region into the low-$\beta$ region in all simulations. In cases (a) and (b) the current extends perpendicular to the direction of the driving motion, which is also the weak field direction for case (b), while in case (c) there is a competition between the driving and weak field directions. However for this moderate value of $k = 0.35$, the extension is not very pronounced.
It is already known that when the null point is asymmetric, the current tends to spread preferentially along the weak field direction in the fan plane, at least when the driving has a non-zero component perpendicular to this direction \citep{alhachami2010,galsgaard2011b}, since the weak field in this region is less able to withstand the field collapse. 
Note that one can in general expect currents to extend along this weak field direction also due to the strong $Q$ values further from the separatrix there, which reflect the variety of field line connectivities present nearby. Such field lines are anchored in distant locations at the boundary, so they typically experience different magnetic stress from the boundary motions, so different perturbed $\BB$. However, at this time in our simulations there has not been time for any communication with the $x$ and $y$ boundaries, and so the current accumulation is associated only with stresses being applied from the spine boundaries and the local collapse dynamics.

\subsection{Predicting flare ribbon locations}
A proper diagnosis of expected locations of flare ribbons would require a self-consistent modelling of particle acceleration in a null point current sheet. This is yet to be done -- most existing studies use simplified analytical models that do not properly represent the structure of the magnetic field and current layer. {We emphasise then that the following analysis based on the MHD approximation is a crude first step toward predicting the expected location of energetic particles.}

\begin{figure}[t]
\centering
\includegraphics[width=0.55\textwidth]{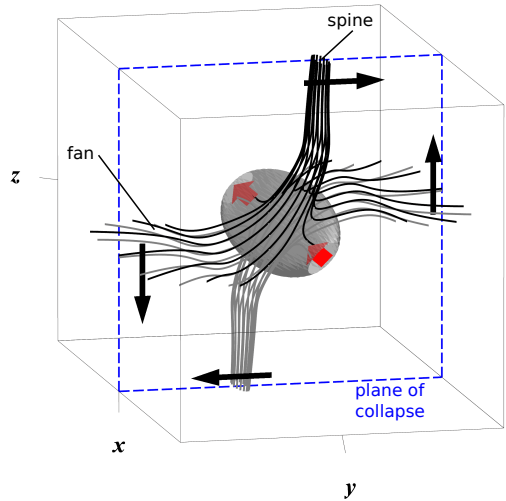}
\caption{Schematic of the structure around a null point at which spine-fan reconnection is taking place. Grey and black lines are magnetic field lines that show the local field collapse. Black arrows indicate the plasma flow. The shaded grey surface outlines the current layer locally along the fan (as in Figure \ref{jsim}), while the red arrows within it indicate the dominant current vector orientation. Modified from \cite{pontin2011b}.}
 \label{spfan}
\end{figure}

First, we should understand the mechanisms by which the acceleration could occur. Null points have been proposed as efficient particle acceleration sites first because the geometry of the field around the null naturally allows for magnetic mirroring, and particle acceleration by gradient-${\bf B}$ and curvature drifts \citep{vekstein1997,petkaki2007,guo2010,stanier2012}. The particle dynamics in the vicinity of the null  may indeed be inherently chaotic \citep{martin1986}. 
In addition, when a current layer is present at the null during reconnection, there can be direct acceleration by the associated electric field. This was observed to be the dominant acceleration mechanism in the PIC simulations of \cite{baumann2013b}, who studied  null point reconnection in the corona using a configuration similar to that of \cite{masson2009}. To understand the resulting particle deposition patterns, one must first understand the structure of the electric current layer at the null.
This electric current distribution is determined in general by a combination of factors. During {\it spine-fan reconnection}, the spine and fan of the null point locally collapse towards one another (see Figure \ref{spfan}) as in our simulations. This collapse of the null occurs in general when a shear perturbation of either the spine or the fan occurs \citep{pontinbhat2007a}. The plane in which this collapse occurs {(plane that contains the deformed spine line -- see Figure \ref{spfan})} is determined both by the perturbation that drives the collapse and the null point structure. The associated current sheet that forms has a current vector that at the null is oriented perpendicular to the plane of collapse, see Figure \ref{spfan}. Thus (in resistive MHD) the parallel electric field is oriented along the fan surface, perpendicular to {the spine and the} plane of the null point collapse \citep{pontinbhat2007a}. Hence, we expect a strong acceleration layer near the null in the fan plane, and thus particle deposition in the vicinity of the fan surface footprint (of oppositely charged particles on opposite sides). It is also possible that particles accelerated {\it towards} the null in this layer may follow the field lines out along the spines. However, when they reach the vicinity of the null point they become effectively de-magnetised, and most particles are simply accelerated `across' the current layer, and out along the fan (rather than being deflected up the spine). Exactly this effect was observed by \cite{baumann2013b}, who noticed very few particles accelerated out along the spines.

We should note that there are at least three factors that could cause enhanced acceleration along the spine structures as well, to create the spine footpoint ribbons observed by \cite{masson2009} and others. 
First, at solar parameters the reconnection process around the site of the original null is likely to be significantly more complex than in the simple models where a single laminar current layer is present. Indeed  this current sheet is susceptible to a tearing-type instability that leads to a fragmented current layer containing many nulls, as described by \cite{wyper2014a}. In such a configuration, the vicinity of the original null becomes highly turbulent, and one would expect efficient particle scattering along both the large-scale spine and fan directions. Second, if there is some large-scale rotational external motion, this can drive {\it torsional spine reconnection}, associated with a component of current parallel to the spine \citep{pontingalsgaard2007} which can accelerate particles along the spine \citep{hosseinpour2014}. Finally, one could expect strong mirroring of particles close to the fan footpoints to lead to a distribution of particles also around the spine footpoints (note that the PIC simulations of \cite{baumann2013b} did not cover the domain all the way to the photosphere).

Based on the above considerations, there exists no unequivocal way of diagnosing general expected particle deposition footprints. However, independent of the details of the acceleration mechanism, if the acceleration happens during reconnection at the null point, then as particles move away from the null they will become magnetised, and would be expected to be observed in the vicinity of either the spine or fan footpoints. For our simulations we make the following basic assumption: Particles will be accelerated in some manner in the vicinity of the current sheet around the null. Thus, we estimate expected deposition patterns by tracing field lines from the current sheet to find their intersections with the boundaries.
To predict particle deposition locations on the spine boundaries ($z=\pm0.5$), we therefore perform the following procedure: We select an array of points that lie on a given current contour level within the domain, and trace field lines from each of these points to the spine boundaries. We then compare the intersections of these field lines with the $Q_\perp$ distribution on the boundary to determine whether they match, i.e.~whether the $Q_\perp$ distribution can be expected to give a good prediction of the geometry of particle deposition signatures. 

\begin{figure}[!t]
\centering
(a)\includegraphics[width=0.45\textwidth]{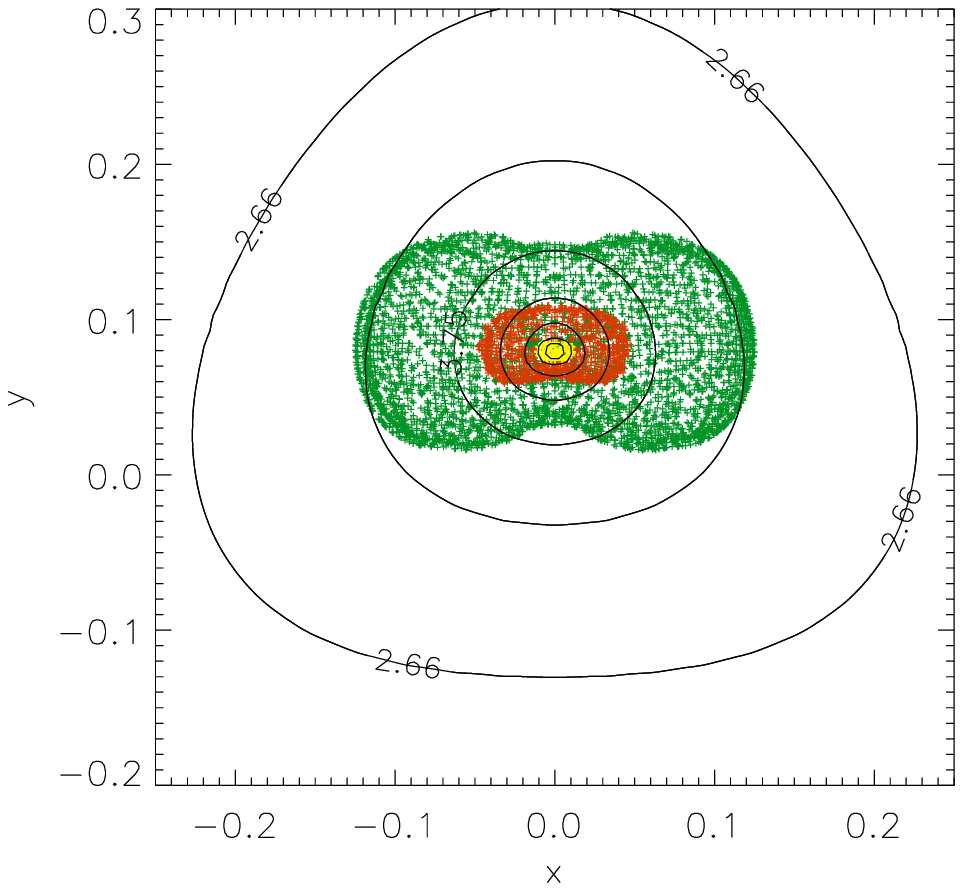}
(b)\includegraphics[width=0.45\textwidth]{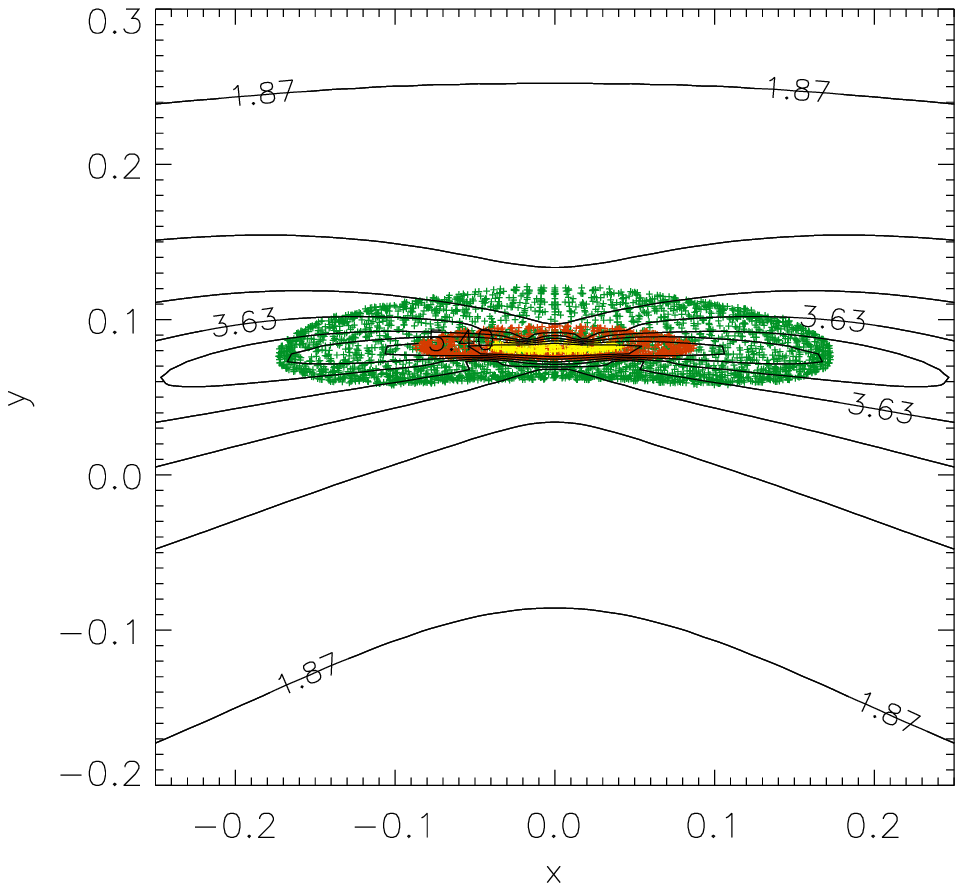}
(c)\includegraphics[width=0.45\textwidth]{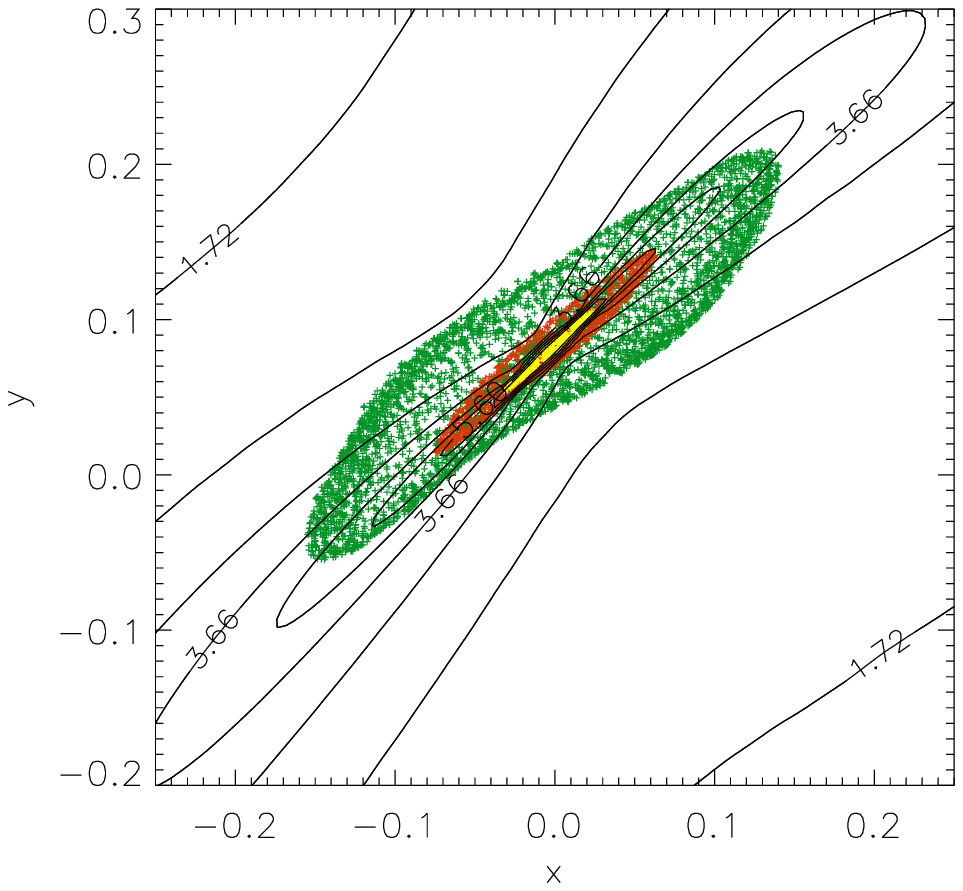}
\caption{$z=0.5$ boundary. 
Coloured crosses show intersection points with this plane of field lines that are traced from initial points lying on a 3D contour of $|\JJ|$ at $t=3.0$.
Selected contours are at 25\%, 50\% and 75\% of the domain maximum (green, red, and yellow crosses, respectively) -- for (a) $k=0.5, \phi=0$, (b) $k=0.35, \phi=0$, (c) $k=0.35,\phi=\pi/4$. Black contours denote $\log_{10}Q_\perp$ as calculated between this plane and a circular cylinder of radius 2.8.}
 \label{jprojspine}
\end{figure}

We first evaluate $Q_\perp$ in our simulations using launch boundary $z=0.5$ and target boundary a circular cylinder of radius 2.8, as in Section \ref{cylinder}.  These contours are plotted, together with footpoint locations for field lines threading the current layer,
 in Fig.~\ref{jprojspine}. Consider first simulation 1 where the field is initially rotationally symmetric (Figs.~\ref{jprojspine}a, \ref{jsim}a). We see that there is some $x$-$y$ asymmetry in the 3D current distribution (Fig.~\ref{jsim}a) that is due to the orientation of the boundary driver. Here the null point field was initially rotationally symmetric, and when we map field lines from the current layer to the boundary we see a similar degree of asymmetry in the 3D current layer and its 2D projection on the boundary (Fig.~\ref{jprojspine}a). 
As a contrast, now compare Figs.~\ref{jsim}(b), \ref{jprojspine}(b). There is little difference in the asymmetry of the 3D current density distribution from Fig.~\ref{jsim}(a). However, the field geometry means that field lines approach the spine more slowly along the weak field direction, so that the projection of $|\JJ|$ is elongated along this direction. This reinforces the fact that the current preferentially spreads along that direction -- but it is the 3D field line geometry rather than the current layer geometry that has the major effect on the projected $|\JJ|$ map. This projection of $|\JJ|$ along the field lines now has a comparable geometry to the $Q_\perp$ contours, although the projected $|\JJ|$ map does not exhibit such high eccentricity/asymmetry. We observe a similar behaviour when we examine Figure \ref{jprojspine}(c), except that in this case (where the boundary driving is at a finite angle to the null point eigenvectors)
the two distributions appear to be rotated with respect to one another -- more pronounced for low current contour levels. These results indicate that a complicated combination of the driving geometry, the field geometry, and the current intensity in the current sheet (itself dependent on plasma parameters and the driving of the system) will influence the expected particle precipitation locations (even using this simple estimate for these locations).

Now consider the boundaries intersected by fan field lines. {The same method as applied before using field line mapping is not useful in determining the angular distribution of expected particle locations, since by definition every field line of the fan connects back to the null and therefore the maximum current region. What is more important for this angular distribution is the orientation of the electric field in the acceleration region together with the global field structure.}
Let us make the following simple considerations.
Particle acceleration along fan field lines will occur through direct acceleration by the DC electric field (or other mechanisms as mentioned above). It is expected that at the null this electric field is directed predominantly towards the weak field direction (perpendicular to the plane of null collapse, see above). However,  the field lines diverge away from this direction, and converge towards the strong field region. Thus the particles {-- which becomes re-magnetised as they are accelerated away from the null --  }are naturally channelled along the field lines into the {neighbourhood of the} strong field direction. This is the region in which $Q$ (or $Q_\perp$) falls off more quickly away from the separatrix -- and thus we expect that particles accumulate around the strong field regions of the fan footprint, corresponding to the locations where the $Q$ contours are narrowest about the fan.

\section{Effect of the global field}\label{globalsec}
\begin{figure}[!t]
\centering
(a)\includegraphics[width=0.5\textwidth]{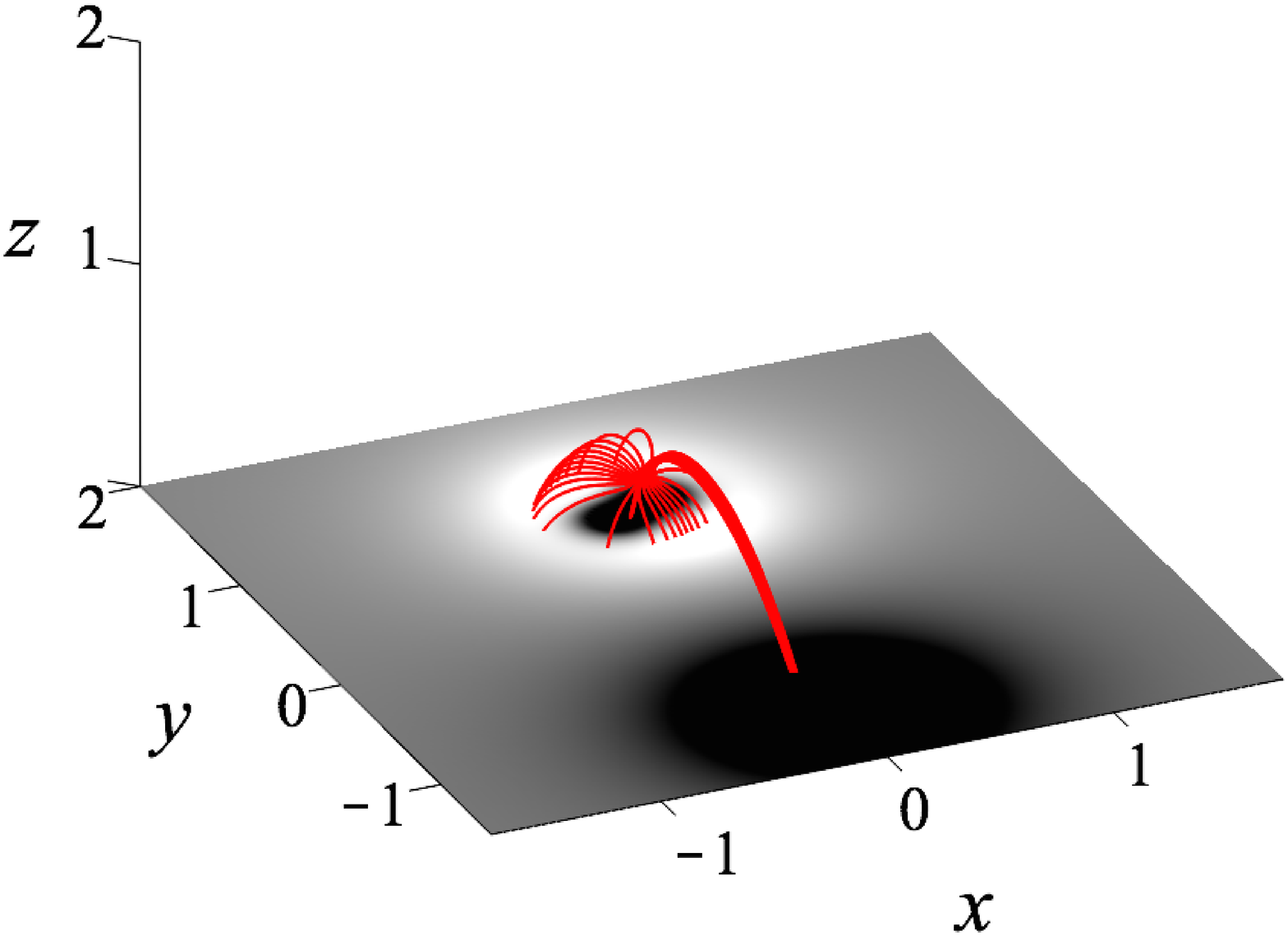}
(b)\includegraphics[width=0.4\textwidth]{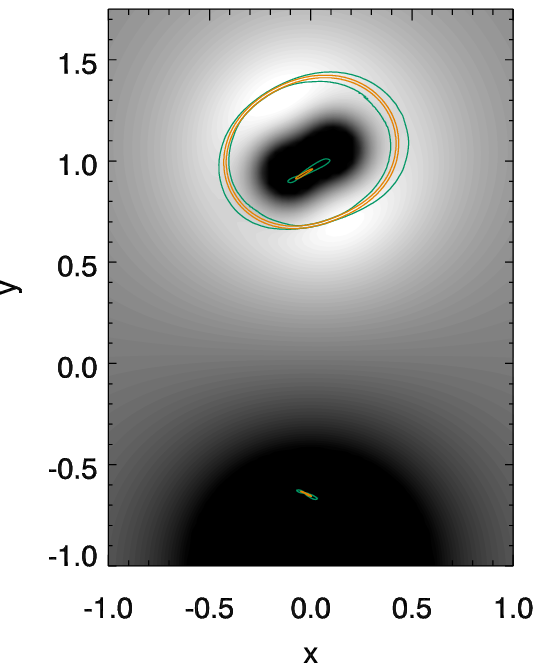}
\caption{(a) Magnetic field lines (red) outlining the spine and fan structure of the coronal null point in the model field of Equation (\ref{b_pot}) with $S=0.109$ (corresponding to $k=0.35$). The shading on $z=0$ shows $B_z$ there, saturated to values $-1$ (black) and +1 (white). (b) $B_z$ distribution on $z=0$ together with contours of the $Q_\perp$ distribution at levels $\log_{10}Q_\perp=\{1.5, 4\}$, coloured green and orange, respectively.}
 \label{domeblines}
\end{figure}

The above sections showed that the asymmetry of the field in the local vicinity of the null can have a profound effect on $Q$, and on the mapping of field lines from the current layer. However, other features of the global field can clearly distort this picture. In this section we return to an equilibrium field and examine the effect of the global coronal geometry, to determine whether the above results regarding the $Q_\perp$-distribution asymmetry carry through beyond the linear null point field. We consider a null point in a separatrix dome configuration, as shown in Figure \ref{domeblines}(a).
The magnetic field is potential, and on the photosphere corresponds to a magnetic dipole, one polarity of which contains an embedded `parasitic polarity'. This field is constructed by placing four magnetic point charges at locations outwith our domain of interest. Specifically, we restrict our studies to the half-space $z>0$, where $z=0$ represents the photosphere, and place all point charges at $z<0$. The magnetic field is given by 
\begin{equation}\label{b_pot}
\BB=\sum_{i=1}^4 \epsilon_i \frac{\xx-\xx_i}{|\xx-\xx_i|^3}
\end{equation}
where $\xx_i$ are the locations and $\epsilon_i$ are the strengths of the point charges. Here we take $\{\epsilon_1,\epsilon_2,\epsilon_3,\epsilon_4\}=\{1,-1,-0.2,-0.2\}$ and $\xx_1=(0,1,-0.5)$, $\xx_2=(0,-1,-0.5)$, $\xx_3=(S,1+S/2,-0.2)$, $\xx_4=(-S,1-S/2,-0.2)$. The charges located at $\xx_3$ and $\xx_4$ are associated with the parasitic polarity around $(x,y)=(0,1)$. The parameter $S$ controls the separation of these two charges. When $S=0$ they are coincident, and the parasitic polarity is approximately circular. However, as $S$ is increased, the parasitic polarity becomes increasingly stretched (see Figure \ref{domeblines}). This elongation of the parasitic polarity means that the field strength becomes less homogeneous in the vicinity of the fan footprint, and also that the field asymmetry at the null -- as measured by the fan plane eigenvalues -- increases. Here we analyse the magnetic topology for three different values of $S$: $S=0$ leads to an approximately symmetric field in the local vicinity of the null that corresponds closely to the symmetric case $k=0.5$ for the linear field, $S=0.067$ that gives fan eigenvalues that correspond to $k=0.45$ for the linear null, and $S=0.109$ that corresponds to $k=0.35$.
\begin{figure}[!t]
\centering
\includegraphics[width=0.9\textwidth]{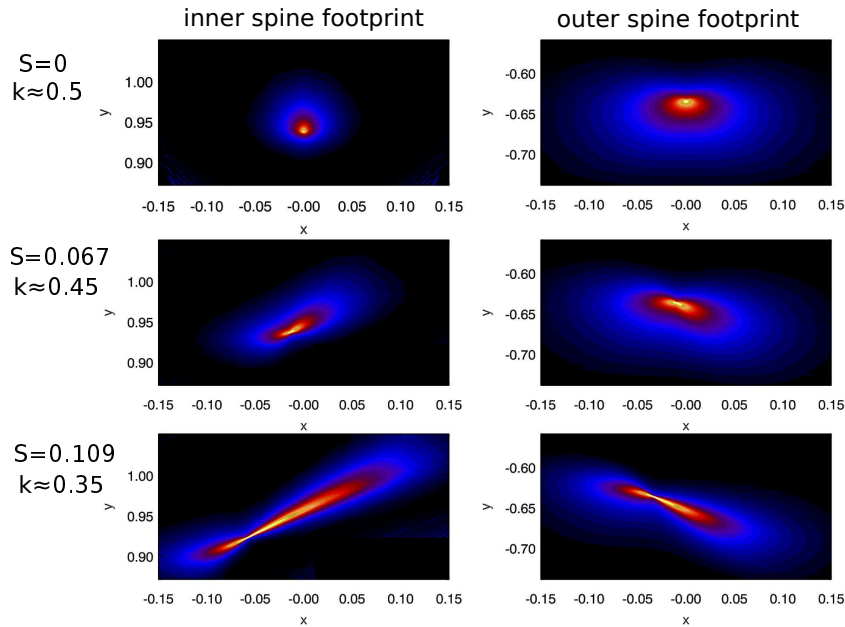}
\caption{Distribution of $Q_\perp$ around the inner (left) and outer (right) spine footprints, for the magnetic field of equation (\ref{b_pot}) with $S=0$ (top), $S=0.067$ (middle), and $S=0.109$ (bottom).}
 \label{domeq}
\end{figure}

As shown in Figure \ref{domeblines}(b), when the null point is asymmetric the level curves of $Q_\perp$ on the photosphere form extended structures. As $S$ is increased, so too the asymmetry of these level curves increases -- both around the inner and outer spine footpoints -- see Figure \ref{domeq}. To determine whether the same scaling with the null asymmetry as above is observed we measure the aspect ratio of these ribbons, defined as follows. The `length' of a given $Q_\perp$ contour is defined as the maximum extent along any line passing through the spine line footpoint (location at which $Q_\perp\to\infty$). The width is then defined as the maximum extent of the contour along any line perpendicular to this, analogous to section \ref{cylinder}. The aspect ratio, being the ratio of the length over the width, is calculated for different $Q_\perp$ contours for each value of $S$, and the results are plotted in Figure \ref{qperp_contours_aspect}. What we see is that there are clear differences in the values of the calculated aspect ratio, both between the inner and outer spine footpoints, and to the results for the linear field (crosses, squares, and circles in Figure \ref{qperp_contours_aspect}, respectively). However, as shown in Eq.~(\ref{aspectratio}) the exact value of the aspect ratio depends on the locations of the launch and target footpoints (for the linear null $a$ and $b$), thus we would not expect an exact agreement.
Moreover, we note that the overall scaling of the aspect ratio with the $Q_\perp$ level is rather well reproduced between the linear field and coronal null point field, and that for the higher $Q_\perp$ contour levels considered the aspect ratios are in rather good agreement.

\section{Discussion and conclusions}\label{discuss}
Coronal magnetic null points exist in abundance, as demonstrated by extrapolations of the coronal field, and have been inferred to be important for a broad range of energetic events. These null points and their associated separatrix and spine field lines are preferential locations for reconnection due to the discontinuity of the field line mapping. This field line mapping also exhibits strong gradients adjacent to the separatrix and spine field lines, that we have analysed here using the squashing factor $Q$ (and $Q_\perp$). Understanding the distribution of $Q$ in the presence of separatrices is of timely importance due to the increasing use of calculation of $Q$ maps in analysing the coronal field topology. 
While a map of the $Q$ distribution shows the presence of both true separatrices and finite-$Q$ QSLs, one should note that 
the physics of current layer formation / energy storage is critically different between a high-$Q$ region containing a separatrix and one that does not. In particular, current singularities are known to form in the ideal limit in the presence of separatrices \citep{pontincraig2005}. Thus reconnection onset is inevitable irrespective of the dissipation (though may be `slow' in an energy storage phase). By contrast, the current layers that form at QSLs are probably finite \citep{craig2014,effenberger2016}, with the onset of reconnection at coronal parameters then requiring a thinning of the QSL and current layer during the energy storage phase \citep{aulanier2005,demoulin2006}. What is clear is that in the case of both null points (separatrices) and QSLs, the current layer formation and eventual dynamics are crucially dependent on the driving of the system, for example from the photosphere.

In this paper, we have made a detailed analysis of the distribution of $Q$ in the presence of magnetic nulls and their associated separatrices. The main results can be summarised as follows.
\begin{enumerate}
\item
It is generically the case that $Q$ is not uniformly distributed around the spine and fan footpoints. Specifically, a generic null point is not rotationally symmetric, and while $Q$ is infinite formally on both the spine and fan of the null, it decays most rapidly away from the spine/fan in the direction in which $|\BB|$ increases most rapidly. When a linearisation of the null is performed (this linearisation characterising the local topology of the field for any topologically stable null \citep{hornig1996}), this direction corresponds to the eigenvector of the largest fan eigenvalue. 
\item
The result of the above is that contours of $Q$ are broadest along the direction of the eigenvalue with smallest fan eigenvalue -- denoted herein as the `weak field direction'. In particular, this demonstrates that the extended, elliptical-like high-$Q$ halo around the spine footpoints observed by, e.g., \cite{masson2009,sun2013} is not a special feature of the particular observations, but is {\it a generic feature when a coronal null is present} whose fan eigenvalues are not equal (i.e.~when the field strength is not homogeneous around the fan footprint).
\item
The asymmetry of the halo of $Q$ contours around the spine/fan increases as the null point asymmetry (measured by the ratio of the eigenvalues) increases. Furthermore, for a given null point asymmetry, the stretching of the $Q$ contours is most extreme for the highest contour levels. 
\item
When the global field geometry (beyond the linear field region) is considered, the  exact aspect ratios of the $Q$ contours are modified from the simple linear null case, but the core of the distribution of $Q$ still reflects the conditions around the null. This is especially true for high $Q$-contour levels.
\item
As a first approximation for understanding why the geometry of flare ribbons is observed to agree well with the geometry of the $Q$ halo in circular ribbon flares \citep[e.g.][]{masson2009}, we analysed MHD simulations of null point reconnection. We traced field lines through the current layer, and analysed the relationship between their intersections with the boundary and the $Q$ contours on the boundary. While no on-to-one relation was found, we showed that field lines traced from the core of the current layer match rather well with the highest $Q$ contours. Thus, particularly for the kernels of the flare ribbons, the $Q$ distribution should in general be expected to predict well the location and orientation of the ribbons. 
\end{enumerate}
It is well established that  an understanding of the null point structure and its relation to the driving of the system is crucial for determining the current layer formation at the null and associated dynamics. We have shown here that this null point structure, defined by its local eigenvectors and eigenvalues,   is intrinsically linked to the distribution of $Q$ away from the spine/fan. Furthermore, this extension of the $Q$ halos around the spine/fan footpoints is in general important for diagnosing the regions of the photosphere that are magnetically connected to any current layer that forms at the null. If we hypothesise this current layer to be a primary site of particle acceleration, this provides predictive properties for e.g.~flare ribbon formation. 
We conclude that the physics in the vicinity of the null and how this is related to the extension of $Q$ away from the spine/fan can be used in tandem to  understand observational signatures of reconnection at coronal null points.

\begin{acks}
The authors acknowledge fruitful discussions with G.~Valori, E.~Pariat, P.~Wyper, E.~Priest and M.~Janvier. D.P. is grateful for financial support from the Leverhulme Trust and the UK's STFC (grant number ST/K000993). The work by K.G. was supported by a research grant (VKR023406) from VILLUM FONDEN.
\end{acks}

\appendix

\section{Calculation of $Q$ and $Q_\perp$ for the linear null}\label{app}
For the linear null point magnetic field of Equation (\ref{beq}), the field line equations ${\rm d}\XX(s)/{\rm d}s=\BB(\XX(s))$ may be solved to obtain parametric equations $\XX(s)$ for the field lines;
\begin{equation}
X(s)=X_0\,{\rm e}^{k s}, \quad Y(s)=Y_0\,{\rm e}^{(1-k) s}, \quad Z(s)=Z_0\,{\rm e}^{-s},
\end{equation}
where $\XX(s)=(X(s),Y(s),Z(s))=(X_0,Y_0,Z_0)$ at $s=0$. Now, set $s=0$ on the plane $z=Z_0=a$, intersecting the spine, so that $(X_0,Y_0,Z_0)=(x,y,a)$. Then we can eliminate $s$ in the above equations, to obtain
\begin{equation}
Y=y\left(\frac{x}{X}\right)^{(k-1)/k},\quad Z=a\left(\frac{x}{X}\right)^{1/k}.
\end{equation}
We now choose the `target plane' to be $(X,Y,Z)=(b,Y,Z)$, $b$ constant. Finally, identifying $\{U,V,u,v\}$ in Eq.~(\ref{qeq}) with $\{Y,Z,x,y\}$, the required derivatives may be obtained. A little algebra leads to the following expression for $Q$:
\begin{equation}
Q(x,y)=\left( {\frac {{y}^{2} \left( k-1 \right) ^{2}}{{k}^{2}{ 
x}^{2}} \left( {\frac {x}{b}} \right) ^{2-2/k}}
+ \left( {\frac {x}{b}} \right) ^{2-2/k}+{\frac {
{a}^{2}}{{k}^{2}{x}^{2}} \left( {\frac {x}{b}} \right) ^
{2/k}} \right)  {\frac {bk}{a}},
\end{equation}
which reduces to Equation (\ref{q_cart}) for $y=0$.

Evaluation of $Q_\perp$ for the same planar boundaries as above requires that we calculate
\begin{equation}
{Q_\perp}^{2}={\frac {{{\partial U}}^{i}}{{{\partial u}
}^{k}} \left( \delta_{{{\it ij}}}-{\frac {
B_{{i}}^\star   B_{{j}}^\star }{ \left| 
\BB^\star  \right|   ^{2}}} \right)  
 \frac{{\partial U}^j}{{\partial u}^l}
 \left( {\delta}^{{
\it lk}}+{\frac {{B}^{l}{B}^{k}}{{B_{{n}}}^{2}}} \right) } \cdot \frac{|\BB^\star|}{|\BB|}
\end{equation}
where $\delta$ is the Kronecker delta, and summation over repeated indices is assumed \citep{titov2007}. In addition, $\{U^1,U^2\}=\{Y,Z\}$, $\{u^1,u^2\}=\{x,y\}$, $B_1^\star,B_2^\star,|\BB^\star|$ are $B_y, B_z, |\BB|$ evaluated at the target boundary $X=b$, and $B^1,B^2,B_n$ are $B_x,B_y,B_z$ evaluated at the launch plane $z=a$. The resulting expression is too lengthy to reproduce here, but reduces to Equation (\ref{qperp_cart}) for $y=0$.


\end{article} 

\end{document}